# Super-resolution in turbulent videos: making profit from damage

**L. Yaroslavsky[*], B. Fishbain, G. Shabat, I. Ideses**

*Department of Interdisciplinary Studies, Faculty of Engineering , Tel Aviv University,*

*Tel Aviv 69978, Israel*

[*]*Corresponding author: yaro@eng.tau.ac.il*

It is shown that one can make use of local instabilities in turbulent video frames to enhance image resolution beyond the limit defined by the image sampling rate. The paper outlines the processing algorithm, presents its experimental verification on simulated and real-life videos and discusses its potentials and limitations.

In long distance observation systems, images and video are frequently damaged by atmospheric turbulence, which causes spatially and temporally chaotic fluctuations in the index of refraction of the atmosphere ([1]) and results in chaotic, spatial and temporal geometrical distortions of neighborhoods of all pixels. This geometrical instability of image frames heavily worsens the quality of videos and hampers their visual analysis. To make visual analysis possible, it is required first of all to stabilize images of stable scenes while preserving real motion of moving objects that might be present in the scene. Methods of generating stabilized videos from turbulent videos, including real time ones, were reported in Refs.[2-4].

In Refs. [3-6], the idea was advanced of making a profit from atmosphere turbulence-induced image geometrical spatial/temporal degradations to compensate image sampling artifacts and generate stabilized images of the stable scene with higher resolution than that defined by the camera sampling grid. In this paper, we elaborate this idea, describe a practical method for producing good quality higher-resolution videos from low-resolution turbulent video streams that implement this idea and illustrate its performance experimentally using simulated and real-time atmosphere distorted videos.

The core of the method is elastic registration, with sub-pixel accuracy, of available video frames of stable scenes followed by resampling of the frames according to the registration results. The first step of the processing of each current frame is updating a reference frame used for elastic registration. For generating and updating the reference frame, temporal pixel-wise median over a block of frames preceding and following the current frame is used ([2-5]).

The next step is computing, with sub-pixel accuracy, a map of pixel displacements in the current frame with respect to the reference frame. It is carried out by means of elastic registration of the current frame and the reference frame. For elastic image registration, several options exist:

(i) optical flow methods ([7,8]); (ii) correlation methods ([9]), (iii) using motion vector data available in MPEG encoded videos ([10]). In the reported experiments, optical flow methods were used.

The pixel displacement map is then analyzed and segmented to separate pixels of the real moving objects from those that belong to the stable scene and are displaced solely due to the atmosphere turbulence. To this goal, the absolute difference from the reference as well as statistical analysis of displacement magnitudes and angles is used ([11]).

The displacement map found for pixels of the stable scene serves, at the next processing step, for placing pixels of each current frame, according to their positions determined by the displacement map, into the reference frame, which is correspondingly up-sampled to match the sub-pixel accuracy of the displacement map. For up-sampling, different image interpolation methods can be used. Among them, discrete sinc-interpolation is the most appropriate as the one with the least interpolation error ([9]). As a result, output stabilized and enhanced in its resolution frame is accumulated. In this accumulation process it may happen that several pixels of different frames are to be placed in the same position in the output enhanced frame. In order to make best use of all of them, these pixels must be averaged. For this averaging, we suggest, computing median of those pixels in order to avoid influence of outliers that may appear due to possible anomalous errors in the displacement map.

After all available input frames of the stable scene are used in this way, the enhanced and up-sampled output frame contains accumulated pixels of input frames in positions, where there were substitutions from input frames, and interpolated pixels of the reference frame in positions, where there were no substitutions,. Substituted pixels introduce to the output frame high frequencies outside the base-band defined by the original sampling rate of the input frames.

Those frequencies were lost in the input frames due to the sampling aliasing effects. Interpolated pixels that were not replaced do not contain frequencies outside the base-band. In order to finalize the processing and make full advantage of the super-resolution provided by the replaced pixels, we suggest the following iterative re-interpolation algorithm that implements discrete sinc-interpolation of non-uniformly sampled data. The algorithm assumes that all accumulated replaced pixels are stored in an auxiliary replacement map that contains pixels values and coordinates. At each iteration of the algorithm, DFT (or DCT, which less vulnerable to boundary effects [19]) spectrum of the image obtained at the previous iteration is computed and then zeroed in all its components outside the selected enhanced bandwidth, say, double the original base-band one. After this, inverse DFT (DCT) is performed of such a modified spectrum and corresponding pixels in the resulted image are replaced by pixels from the replacement map thus producing an image for the next iteration. In this process, the energy of the zeroed outside spectrum components can be used as an indicator when the iterations can be stopped.

Stabilized and resolution enhanced images, obtained in this way can be finally down-sampled to the sampling rate determined by desired enhanced bandwidth and then subjected to additional processing aimed at camera aperture correction and, when necessary, denoising.

The efficiency of the entire processing is illustrated in Fig. 1 by the results of computer simulations and in Figs.2 for a real-life video. As one can see from Figure 1, even from as small number of the frames as 15, a substantial resolution enhancement is possible especially when a sufficiently large number of re-interpolation iterations are carried out. Fig. 2, a) and (b) shows 256x256 pixel fragment of a stabilized frame of real-life obtained as a temporal median over 117 frames, and the same fragment obtained after replacing its pixels, as it is described above, by pixels of those 117 frames according to their displacement maps. As visually resolution improvement can be appreciated only on a high quality display, image c) presents difference between these two images that clearly shows edges enhanced in image (b) compared with image (a). Fig. 2 d) presents the final result of the processing after re-interpolation and aperture correction implemented in the assumption that camera sensor has fill-factor close to 1.

For quantitative evaluation of the resolution improvement we used a method for numerical evaluation of image visual quality degradation due to blur suggested in [12]. The method is based on the discrimination between different levels of blur perceptible on the same picture. The image degradation measure ranges from 0 to 1, which are respectively the lowest and the highest degradations. Degradation factors for images shown in Fig. 2 are 0.43 for the stabilized image before resolution enhancement, 0.4 for resolution-enhanced image and 0.28 for the final re-interpolated and aperture corrected image.

In the evaluation of the results obtained for real-life video, one should take into account that substantial resolution enhancement in the described super-resolution fusion process can be expected only if the camera fill-factor is small enough. The camera fill-factor determines the degree of low-pass filtering introduced by the camera. Due to this low pass filtering, image high frequencies in the base-band and aliasing high frequency components that come into the base-band due to image sampling are suppressed. Those aliasing components can be recovered and returned back to their true frequencies outside the base-band in the described super-resolution process, but only if they have not been lost due to the camera low pass filtering. The larger fill-factor, the heavier unrecoverable resolution losses. In the described simulation experiment, camera fill-factor is 0.05, whereas in reality fill-factors of monochrome cameras are usually close to 1.

In conclusion, one can state that presented results show that the described technique allows, in addition to compensating atmospheric turbulence in video sequences, to improve image resolution thanks to proper elastic local registration and re-sampling of degraded video frames. The degree of the achievable in this way resolution improvement depends on the number of frames that contain stable scene, spread of turbulence-induced local image displacements and on the camera fill-factor. Certain resolution enhancement can be achieved even for conventional cameras with large fill factor. For cameras with small fill-factor, such as, for instance, color cameras, the potential resolution enhancement might be, as simulation experiments show, very substantial.

References

1. M. C. Roggermann and B. Welsh," Imaging Through Turbulence", CRC Press, Inc, 1996.

2. Shai Gepshtein, Alex Shtainman, Barak Fishbain, Leonid Yaroslavsky, "Restoration of atmospheric turbulent video containing real motion using elastic image registration", *The 2004 European Signal Processing Conference (EUSIPCO-2004)*, John Platt, pp. 477-480, Vienna, Austria, 2004.

3. L. Yaroslavsky, B. Fishbain, I. Ideses, D. Slasky, Z. Hadas, "Simple Methods for Real Time Stabilization of Turbulent Video", *Proc. of ICO Topical Meeting on Optoinformatics/ Information Photonics*, Maria L. Calvo, Alexander V. Pavlov and Jurgen Jahns, ITMO, pp. 138-140, St Petersburg, Russia, 2006.

4. B. Fishbain, L. P. Yaroslavsky, I. A. Ideses, A. Shtern, O. Ben-Zvi, "Real-time stabilization of long-range observation system turbulent video", *Proc. of Real-Time Image Processing/Electronic Imaging 2007*, SPIE Vol. 6496, San-Jose, CA, USA, 28 January – 1 February 2007.

5. B. Fishbain, I.A. Idess, Sh. Gepstein, L.P. Yaroslavsky, Turbulent video enhancement: Image stabilization and super-resolution, *Proc. The 11th. Meeting on Optical Engineering and Science in Israel*, 2007

6. Z. Zalevsky, Sh. Rozental, and M. Meller, Usage of Turbulence for Super Resolved Imaging, *Optics Letters, v. 32, No. 9, May 1, 2007, p. 1072*

7. T. Brox, A. Bruhn, N. Papenberg & J. Weickert, High Accuracy Optical Flow Estimation based on Theory for Wrapping, *Proc. 8th European Conference on Computer Vision*, Prague, Check Republic, 2004, 4:25-36.

8. L. J. Barron, D. J. Fleet and S. S. Beachemin, Performance of Optical Flow Techniques, *International Journal of Computer Vision 12*, 1994, 43-77.

9. L. Yaroslavsky, Digital Holography and Digital Image Processing, Kluwer Scientific Publishers, Boston, 2004

10. I. Ideses, , L. Yaroslavsky, B. Fishbain, R. Vistuch, 3D from compressed 2D video, Stereoscopic Displays and Virtual Reality Systems XIV. Edited by Woods, Andrew J.; Dodgson, Neil A.; Merritt, John O.; Bolas, Mark T.; McDowall, Ian E.. Proceedings of the SPIE, Volume 6490, pp. 64901C (2007)

11. B. Fishbain, L. Yaroslavsky, I. Ideses, O. Ben-Zvi, and A. Shtern , Real time stabilization of long range observation system turbulent video , Proc. SPIE  6496, 64960C (2007)

12. F. Crete, T. Dolmiere, P. Ladret, M. Nicolas, "The Blur Effect: Perception and Estimation with a New No-Reference Perceptual Blur Metric", *Proc. of Real-Time Image Processing/Electronic Imaging 2007*, SPIE Vol. 6492, San-Jose, CA, 2007.

List of Figures

Fig. 1. Illustrative simulation results of resolution enhancement of turbulent video frames: a)- initial high resolution frame; b) - an example of low resolution frame obtained by camera with fill-factor 0.05; c) - resolution enhanced frame obtained by the described fusion process with 50 iterations of the re-interpolation from 15 low resolution frames distorted by simulated random local displacements with standard deviation 0.5 inter-pixel distance.

Fig. 2. Illustration of resolution enhancement of real-life video frames a) – a 256x256 pixels fragment of the stabilized frame; b) the same fragment after resolution enhancement; d) difference between images a) and b) that show edges enhanced in image b); d) image b) after re-interpolation and aperture correction.

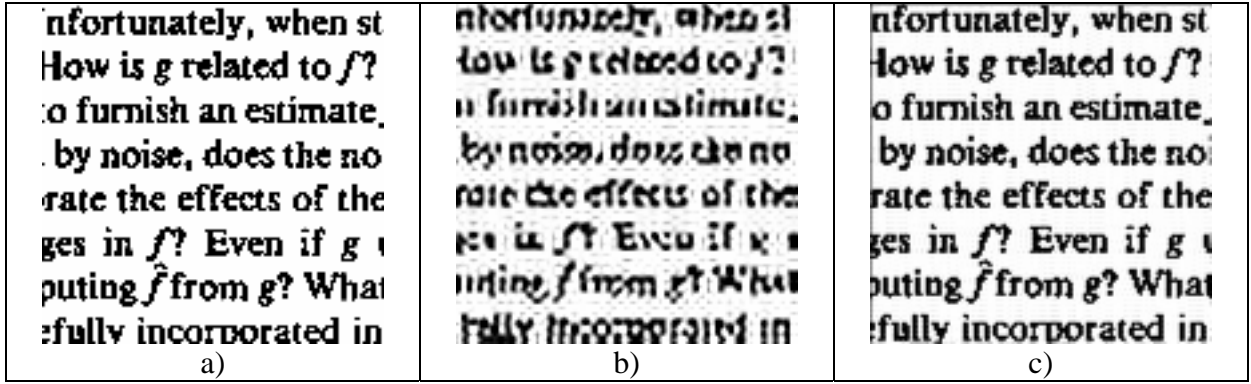

Fig. 1. Illustrative simulation results of resolution enhancement of turbulent video frames. a)- initial high resolution frame; b) - an example of low resolution frame obtained by camera with fill-factor 0.05; c) - resolution enhanced frame obtained by the described fusion process with 50 iterations of the re-interpolation from 15 low resolution frames distorted by simulated random local displacements with standard deviation 0.5 inter-pixel distance.

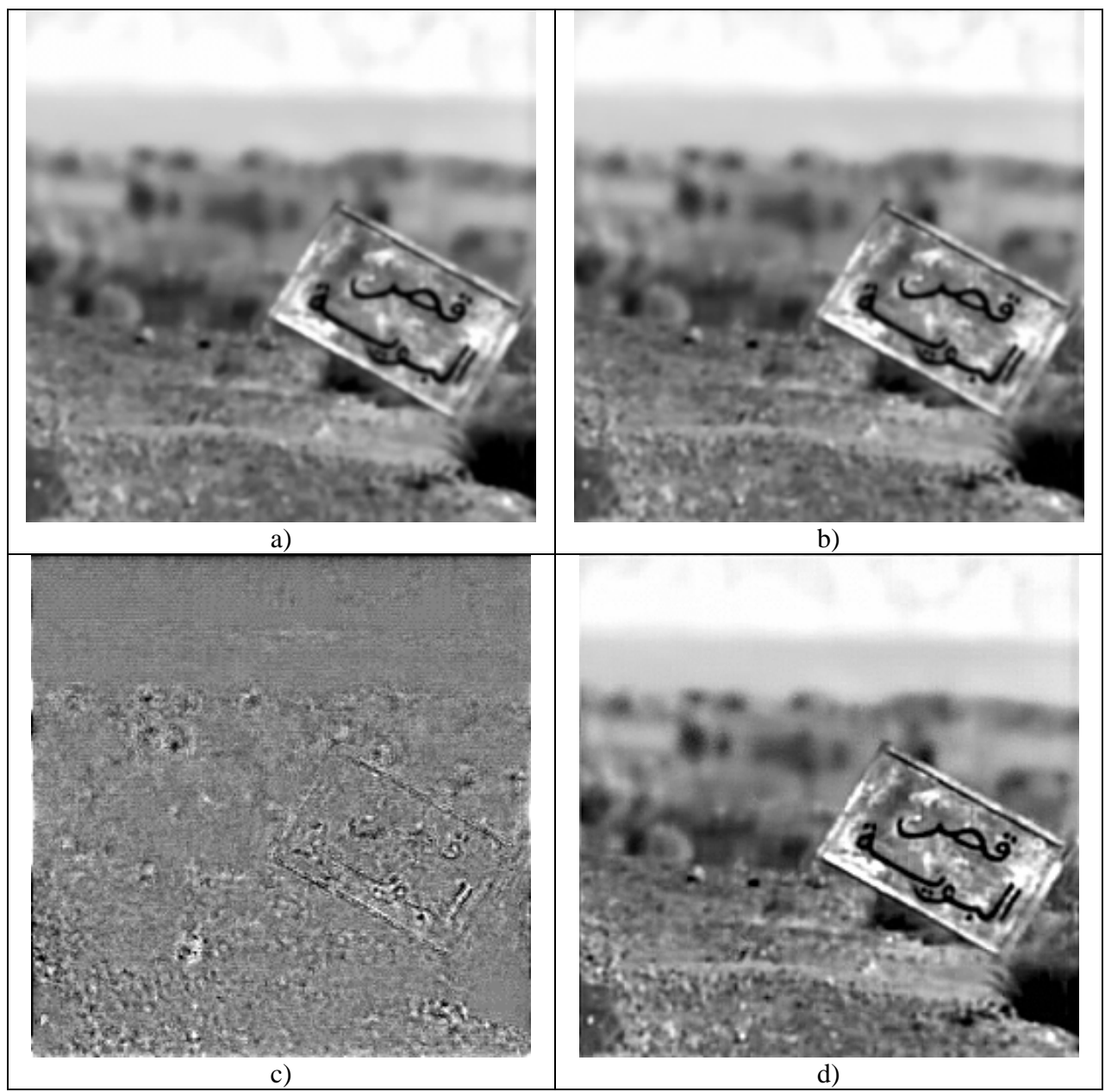

Fig. 2. Illustration of resolution enhancement of real-life video frames a) – a 256x256 pixels fragment of the stabilized frame; b) the same fragment after resolution enhancement; d) difference between images a) and b) that show edges enhanced in image b) as compared to image a); d) image b) after re-interpolation and aperture correction.